\documentclass[10pt,twoside]{article}
\usepackage{graphicx}
\usepackage{latexsym}
\usepackage{amsfonts}
\usepackage{amssymb}
\usepackage{amsmath}

\makeindex

\begin{document}

\title{On The Higher Codimension Braneworld}

\markboth{Sugumi Kanno, Jiro Soda}{On The Higher Codimension Braneworld}

\author{Sugumi Kanno, Jiro Soda
\\[5mm]
\it Department of Physics,  Kyoto University, Kyoto 606-8501, Japan\\
\it e-mail: sugumi@tap.scphys.kyoto-u.ac.jp, jiro@tap.scphys.kyoto-u.ac.jp
}

\date{}
\maketitle

\thispagestyle{empty}

\begin{abstract}
\noindent
We study a codimension 2  braneworld
 in the Einstein-Gauss-Bonnet gravity.  
 In the linear regime, 
  we show the conventional Einstein gravity can be recovered on the brane.
 While, in the nonlinear regime, we find corrections due to the thickness 
 and the bulk geometry. 

\noindent
{\bf Keywords:} Braneworld, Codimension 2, Thickness
\end{abstract}

\section{Introduction}

 The old idea that our universe may be a braneworld embedded in a 
 higher dimensional spacetime is renewed by the recent development
 in string theory which can be  consistently formulated only in 10 
 dimensions. Instead of 10 dimensions, however, 
 most studies of braneworld cosmology have been 
 devoted to 5-dimensional models.  Needless to say,
 it is important to investigate the higher codimension braneworld. 
   In this paper, we investigate a codimension 2 braneworld 
  in the Einstein-Gauss-Bonnet gravity taking into account the thickness. 
  By examining the structure of the singularity in the equations of motion,
  we find a possibility to treat the thickness within the context of the
  distributional source. We name it a quasi-thick braneworld.

\section{Quasi-thick Braneworld}

We consider a codimension 2 braneworld  with a positive tension in 
the 6-dimensional bulk spacetime described by the action
\begin{eqnarray}
S=\frac{1}{2\kappa^2}\int d^6x\sqrt{-g_{(6)}}
	\left[R + \alpha R^2_{\rm GB}\right]
	-\int d^4x\sqrt{-g}~\sigma 
	+\int d^4x\sqrt{-g}{\cal L}_{\rm matter}  \ ,
\end{eqnarray}
where $\kappa^2$ is the 6-dimensional gravitational constant, 
$g_{(6)\mu\nu}$ and $g_{\mu\nu}$ are the 6-dimensional bulk 
and the our 4-dimensional brane metrics, respectively. 
Here, ${\cal L}_{\rm matter}$ is the Lagrangian density of the matter 
on the brane, and $\sigma$ is the brane tension. 
The Gauss-Bonnet (GB) term is given by
$
R^2_{\rm GB}=R^{abcd}R_{abcd}
	-4R^{ab}R_{ab}+R^2  \ .
	\label{GBterm}
$
The Latin indices $\{a,b,\cdots\}$ and the Greek indices $\{\mu,\nu,\cdots\}$
are used for tensors defined in the bulk and on the brane, respectively.

 We will assume that a 6-dimensional metric has axial symmetry, which reads 
\begin{eqnarray}
ds^2=dr^2+ g_{\mu\nu} (r,x^\mu )  dx^\mu dx^\nu
	+L^2(r,x^\mu)d\theta^2  \ . 
	\label{metric}
\end{eqnarray}
This assumption corresponds to the $Z_2$ symmetry in the 
 Randall-Sundrum braneworld model. 
Here we have introduced polar coordinates $(r,\theta)$ for 
the two extra spatial dimensions, where 
$0\leq r < \infty$ and $0 \leq\theta < 2\pi$. 
As we locate a four-dimensional brane at $r=0$, which is 
a string like defect, we must take the boundary condition
$
 \lim_{r\rightarrow 0} L (r , x^\mu ) = 0 \ ,
 \lim_{r\rightarrow 0} L' (r , x^\mu ) = {\rm const.} 
 \label{bc:L}\ ,
$ 
where the prime denotes derivatives with respect to $r$.
The first condition realizes the 4-dimensional
brane at $r=0$. And the second condition
allows the existence of the conical singularity. 
The structure of conical singularity is a 2-dimensional delta 
function $\delta(r)/L$.

 Possible components of the energy-momentum 
tensor $T_{ab}$, which could be balanced with the singular part of 
 Einstein tensor, are given by
\begin{eqnarray}
T_{ab} ({\rm singular})=
	\left(\begin{array}{ccc}
	0&0&0\\
	0&T_{\mu\nu}\frac{\delta(r)}{2\pi L}+S_{\mu\nu}
	\delta(r)&0\\
	0&0&S_{\theta\theta}\delta(r)\\
	\end{array}\right)  \ ,
	\label{EM}
\end{eqnarray}
where $T_{\mu\nu}$ represents the conventional brane matter, 
while $S_{\mu\nu}$ and $S_{\theta\theta}$ describe
 the extra matter which mimics the thickness of the braneworld.
 In this way, the thickness can be treated  within the context of a 
 distributional source, which we dub quasi-thickness.

\section{Effective Theory}
%
\subsection{Linear Regime}

We now consider a perturbation around the vacuum solution   
\begin{eqnarray}
ds^2=dr^2+ \eta_{\mu\nu}dx^\mu dx^\nu
	+c^2r^2d\theta^2\ ,
\end{eqnarray}
where $c$ is a constant of integration. For $c\ne 1$, we have
a conical singularity at $r=0$.
Then the matching condition determines 
the deficit angle in terms of the brane tension as
$
1-c =\kappa^2 {\sigma / 2\pi} \ .
$
Note that $c < 1$, because the tension $\sigma$ is positive.
Here we will discuss only the scalar perturbation. 
In the case of vector and tensor perturbations, 
see our paper~\cite{Kanno:2004nr}. 
The final result is written by
\begin{eqnarray}
ds^2 =  dr^2
	+\left(1+h_{\mu\nu}\right)
	\eta_{\mu\nu}dx^\mu dx^\nu
	+L^2d\theta^2\ ,
\end{eqnarray}
where
\begin{eqnarray}
h_{\mu\nu}&=&\frac{C_0}{2}+\frac{3}{r}Z_1(r,x^\mu)
	-\frac{3}{2}Z_2(r,x^\mu)
	+\int^r_0dr \int^r_0dr\delta g_{rr ,\mu\nu}
	-2\chi_{,\mu\nu}~r\ ,  \nonumber\\
\delta g_{rr}&=&\frac{C_0}{2}-\frac{3}{r}Z_1(r,x^\mu)
	-\frac{3}{2}Z_2(r,x^\mu)
	\label{scalar:metric}  \ , \\
L^2&=&c^2r^2\left[1-\frac{3}{2}C_0
	-\frac{3}{r}Z_1(r,x^\mu)
	+\frac{9}{2}Z_2(r,x^\mu)\right]
	- c^2 r \int^r_0dr\delta g_{rr}
	+2c^2r\chi\ , \label{scalar:L}
\end{eqnarray}
Here, we have defined
\begin{eqnarray}
Z_1(r,x^\mu) = \int d^4p~e^{ip\cdot x}A(q)J_1(qr)\ , \ 
Z_2(r,x^\mu) = \int d^4p~e^{ip\cdot x}A(q)qJ_2(qr)\ .
\end{eqnarray}
Now, the brane is located at $r=0$ in this coordinate system.
Note that the distortion function $\chi(x^\mu)$ takes into account
the fact that the brane is not necessarily
located at the constant $r$ in the coordinate system 
which we used to solve equations in the bulk.

 Since we have solved the bulk geometry, we next discuss the 
 matching conditions. Using the energy-momentum tensor (\ref{EM}),
matching conditions reduce to the effective theory on the brane.
 Using Eq.~(\ref{scalar:L}),
 the deficit angle can be obtained from
$
 \lim_{\epsilon \rightarrow 0}
  L' (r =\epsilon ) 
 = c(1-C_0)\ .
$
This tells us that the deficit angle is perturbed on the brane 
at the linear level, but it is constant. 
 From Eq.~(\ref{scalar:metric}), the extrinsic curvature 
 in this limit gives 
$
 \lim_{\epsilon \rightarrow 0} 
  K_{\mu\nu}(r=\epsilon,x^\mu) = \chi_{,\mu\nu} \ .
$
From this result,
we see that the extrinsic curvature is nonzero near the brane, 
if the distortion field $\chi (x^\mu)$ exists.
 In the coordinate system where we have solved the equations
 of motion in the bulk, the location of the brane is specified by $\chi (x^\mu)$
 and the resulting shape of the braneworld becomes the deformed cylinder
 in which the distortion is represented by $\chi (x^\mu)$ (see Fig.1). 
 At first sight, this deformed cylinder looks like a 5-dimensional
 hypersurface. However,
 at the location of the brane ($r=\epsilon$ in the Gaussian normal gauge),  
 the circumference radius of the cylinder vanishes because of 
 ${\displaystyle\lim_{\epsilon \rightarrow 0}} L(r=\epsilon ,x^\mu) =0 $ there.
 Hence, the deformed cylinder is the 4-dimensional braneworld. 
\begin{figure}[h]
\centerline{\includegraphics[height=4.5cm, width=4cm]{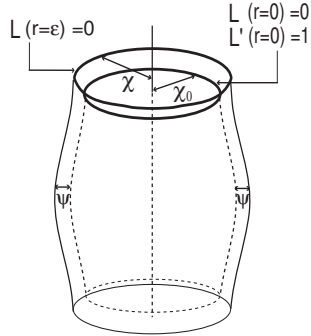}}
\caption{Schematic picture for the quasi-thick braneworld. 
 }
\end{figure}

 Now, we must specify $K_{\mu\nu}(r=0,x^\mu )$ which is not known a priori.
 Without losing the generality, we can write
$
  K_{\mu\nu}(r=0,x^\mu ) = \chi_{0,\mu\nu} (x^\mu ) + \rho_{\mu\nu} (x^\mu )
  \ , \quad \rho^\mu{}_\mu =0 \ , 
$
where we decomposed the extrinsic curvature at $r=0$ into the traceless part
 $\rho_{\mu\nu} (x^\mu)$ and the trace part $ \square \chi_0 (x^\mu )$. 
 As the field $\chi (x^\mu )$ represents the location of the brane 
 $r=\epsilon $, it is natural to regard $\chi_0 (x^\mu )$ 
 as the location of $r=0$. 
 Hence,  $\psi (x^\mu ) \equiv \chi (x^\mu ) -\chi_0 (x^\mu )$ 
 can be interpreted as  the thickness of the braneworld.

The matching condition of ($\mu,\nu$) component leads to
\begin{eqnarray}
G^\mu{}_\nu =\frac{\kappa^2}{8\pi\alpha(1-c)}T^{\mu}{}_{\nu}
	+\delta^\mu_\nu\frac{cC_0}{4\alpha(1-c)}  \ ,
	\label{main}
\end{eqnarray}
 where the second term is the cosmological constant.
 Eq.~(\ref{main}) proves that Einstein gravity is recovered for 
 the codimension 2 braneworld in the case of the linearized gravity.
 This fact  is consistent with the result in~\cite{Bostock}.
Considering the other components of the maching condition and 
conservation law of energy-momentum tensor, we get the relation 
$\chi_0 = c \chi$. Since $c<1$,  $\psi=\chi-\chi_0= (1-c)\chi$ is positive.
 Hence, it is legitimate to interpret $\psi$  as the effective thickness.  
 It should be noted that the effective thickness $\psi$ satisfies
$
\square\psi=-\frac{\kappa^2}{3}S^\mu{}_\mu\ .
$
This is reminiscent of the radion in the codimension 1 
braneworld.

%
\subsection{Nonlinear Regime}

 The importance of the quasi-thickness  can be
 manifest in the next order calculations. 
Up to the second order, we expect 
\begin{eqnarray}
G^\mu{}_\nu&=&\frac{\kappa^2}{8\pi\alpha (1-c+cC_0)}T^\mu{}_\nu 
	+\frac{cC_0-\overset{(2)}{L}{}^{\prime}}{4\alpha (1-c+cC_0)}
	\delta^\mu_\nu
	\nonumber\\
&&
	+ c\Big[\chi^{,\mu}{}_{,\alpha} \chi^{,\alpha}{}_{,\nu}
	- \square \chi \chi^{, \mu}{}_{, \nu}
	+\frac{1}{2} \delta^\mu_\nu \left\{ (\square\chi)^2 
	-\chi^{,\alpha}{}_{,\beta} \chi^{,\beta}{}_{,\alpha}\right\}
	\Big] \nonumber \\
&&	-\frac{1}{1-c} \left[ \rho^\mu{}_\alpha \rho^\alpha{}_\nu
	- \delta^\mu_\nu \rho^\alpha{}_\beta \rho^\beta{}_\alpha
	      \right]   \label{second-order} \ ,
\end{eqnarray}
where the corrections in the last two lines carry the information 
of thickness of the brane. Another correction
come from $\overset{(2)}{L}{}^{\prime}$ carries the effects of
bulk, that is, the deformed deficit angle. This needs the next
order analysis. Thus, it turns out that some corrections due 
to the quasi-thickness can be expected at the second order. 

\section{Conclusion}

We have shown that the Gauss-Bonnet gravity gives a consistent 
framework discussing the quasi-thick codimension 2 braneworld. 

\section*{Acknowledgements}
This work was supported in part by  Grant-in-Aid for  Scientific
Research Fund of the Ministry of Education, Science and Culture of Japan 
 No. 155476 (SK) and  No.14540258 (JS) and also
  by a Grant-in-Aid for the 21st Century COE ``Center for
  Diversity and Universality in Physics".

\end{document}